\documentclass{ws-ijmpa}
\usepackage[super,compress]{cite}
\usepackage{graphicx}
\begin{document}
\markboth{A. Tartaglia}{The strained state cosmology}

%
\catchline{}{}{}{}{}
%

\title{The strained state cosmology}

\author{Angelo Tartaglia}

\address{Department of Applied Science and Technology, Politecnico, Corso Duca degli Abruzzi 24\\
Torino, 10129, Italy\\
angelo.tartaglia@polito.it}

\maketitle

\begin{history}
\received{Day Month Year}
\revised{Day Month Year}
\end{history}

\begin{abstract}
Starting from some relevant facts concerning the behaviour of the universe over large scale and time span, the analogy between the geometric approach of General Relativity and the classical description of an elastic strained material continuum is discussed. Extending the elastic deformation approach to four dimensions it is shown that the accelerated expansion of the universe is recovered. The strain field of space-time reproduces properties similar to the ones ascribed to the dark energy currently called in to explain the accelerated expansion. The strain field in the primordial universe behaves as radiation, but asymptotically it reproduces the cosmological constant. Subjecting the theory to a number of cosmological tests confirms the soundness of the approach and gives an optimal value for the one parameter of the model, i.e. the bulk modulus of the space-time continuum. Finally various aspects of the Strained State Cosmology (SSC) are discussed and contrasted with some non-linear massive gravity theories. The possible role of structure topological defects is also mentioned. The conclusion is that SSC is at least as good as the $\Lambda CDM$ standard cosmology, giving a more intuitive interpretation of the physical nature of the phenomena.

\end{abstract}

\ccode{PACS numbers: 95.30.Sf; 98.80.-k}


\section{Introduction}	

Our access to the knowledge of the physical universe at a large scale is through observation. For a long time the observation instruments have just been our eyes, but today much more sophisticated tools are available, allowing to explore the whole electromagnetic spectrum from the sky and to record and analyze incoming particle fluxes. From all that we acquire abundance of information and identify a number of facts whose origin is scarcely known. Of the latter I quote just three cases:
 \begin{romanlist}
 \item galaxies are unevenly scattered in the sky, giving rise to a three-dimensional spongy pattern, with big voids delimited by walls and filaments;
 \item apparently, 'something' is pushing space to expand, but we do not know what it is; it seems to produce expansion, but no other visible effect; in particular it does not gravitate;
 \item at various scales localized gravitational effects exist, whose source is otherwise invisible and does not coincide with ordinary matter.
\end{romanlist}

Coming to the conceptual framework allowing us to interpret the facts, we know that:
\begin{itemlist}
\item apparently the gravitational interaction is very well described as a geometric property of a four-dimensional Riemannian manifold with Lorentzian signature;
\item the other fundamental interactions do not share this geometric essence.
\end {itemlist}

Our understanding of the physical world encompasses an unresolved conflict, that may be evidenced writing Einstein's equations for the gravitational field, i.e. the core of the General Relativity (GR) theory:
\begin{equation}
G_{\mu\nu} = \kappa T_{\mu\nu}.
\label{einstein}
\end{equation}

On the left the Einstein tensor follows the rules of continuous geometry. On the right, the energy/momentum tensor of matter/energy depends on the properties of quantum operators acting in abstract spaces. Apparently nature is dual: space-time (and continuous geometry) on one side; matter/energy fields (and quantum mechanics) on the other.

The two paradigms conflict with each other and nobody has so far really solved the conflict.
\subsection{Axiomatic assumptions}
The theoretical description of the physical phenomena, besides the rules of logic, currently rests on an axiomatic (i.e. not formally motivated) assumption:
\begin{itemlist}
\item the physical configuration of the world, including all interactions, satisfies an universal 'economy' principle, known as the least (or extremal) action principle.
\end{itemlist}
The mathematical formulation of the 'principle' is $\delta S = 0$, being the action integral $S$ written:
\begin{equation}
S=\int{\ast\Lambda}=\int{\textit{L}d^Nx}.
\label{action}
\end{equation}

In the coordinates-free formulation $\Lambda$ is an N-form (in an N-dimensional manifold) and $\ast$ stands for Hodge conjugation. Once coordinates have been fixed, $d^Nx$ is the N-dimensional volume element and $\textit{L}$ is the Lagrangian density of the system. The integral is over a volume whose boundary is fixed.

The problem with (\ref{action}) is how to choose the Lagrangian density. In the 18th century and for simple mechanical systems it looked reasonable to choose for $\textit{L}$ the difference between the kinetic and the potential energy densities of the system. In general $\textit{L}$ is some function of the coordinates (including time) and their first order derivatives with respect to some affine parameter. Additional heuristic criteria for building a Lagrangian density are:
\begin{itemlist}
\item the Lagrangian density should be 'as simple as possible' (a sort of minimal complexity principle),
\item at least one minimum should exist (which implies that the first order derivatives should be at least power-two terms).
\end{itemlist}
The motivations for such criteria are practical and aesthetical, rather than mandatory, so, often, facing some problem in classical or quantum field theory, people indulge to practicing a sort of 'Lagrangian engineering', releasing the first of the above constraints. The practice is driven by the results one is trying to obtain: is that motivation enough? For sure fancy Lagrangians lack simple physical interpretations.

In the following I shall propose an approach to the description of the properties of space-time based on a simple and intuitive analogy with elastic continua, leading, as I shall show, to a consistent interpretation of the cosmic accelerated expansion.

\section{N-dimensional Analogy between Elasticity and Geometry}

Continuous media are characterized by their intrinsic geometry. When some distortion, due to external or internal causes, is allowed, the intrinsic geometry changes. In the ordinary theory of elasticity, the medium is three-dimensional in space, then time appears as an evolution parameter. This geometrical approach is easily extended to any number $N$ of dimensions and, introducing the appropriate signature in the manifold, time becomes part of the continuum: the description is now relativistic and our deformable medium is space-time as such.

Start from an undeformed N-dimensional continuum: no internal defect, no external agent acting upon the medium. The natural intrinsic geometry of the manifold is either Euclidean or Minkowskian, according to the choice for signature. The typical line element is:
\begin{equation}
dl_0^2=E_{ij}dx^idx^j,
\label{line0}
\end{equation}
where $E_{ij}$ is an element of the Euclidean/Minkowskian metric tensor (in the Euclidean case and for Cartesian coordinates it is $E_{ij}=\delta_{ij}$). The indices $i$ and $j$ range from 1 to N.

Introduce now a distortion, originating from any physical cause. Every little volume in the medium will be displaced to a new position; excluding rigid translation and global rotations, the geometry will change and the new line element will be:
\begin{equation}
dl^2=g_{ij}dx^idx^j,
\label{line1}
\end{equation}
where the new metric tensor is globally different from the previous one that no longer exists: $g_{ij}\neq E_{ij}$.
The tensor which encompasses the distortion of the original manifold is the strain tensor $\epsilon_{ij}$:
\begin{equation}
\epsilon_{ij}=\frac{1}{2}(g_{ij}-E_{ij}).
\label{strain}
\end{equation}

It is important to stress that $E_{ij}$ is used just to build the strain tensor $\epsilon_{ij}$ and has neither the properties nor the role of a metric tensor for the distorted manifold (the only one existing).
In order to move from a purely formal description to physics, I assume that the distortion is \textit{elastic}, i.e. that the underlying interactions within the continuum insure that, removing the cause of the strain, the manifold reverts to the original unstrained state. We know that in the real world this never happens entirely, but may be assumed as an acceptable approximation in most cases.

The next steps may be made adopting a thermodynamical approach, introducing the free energy density, $\mathfrak{F}$, of the continuum and writing it as a power expansion of the scalars that can be built out of the strain tensor, whose components are now the Lagrange coordinates of the system:
\begin{equation}
\mathfrak{F}=\mathfrak{F}_0+\frac{\lambda}{2}(\epsilon_i^i)^2+\mu\epsilon_{ij}\epsilon^{ji}+...,
\label{freeenergy}
\end{equation}
where $\mathfrak{F}_0$ is a constant; the standard tensor notation is used and the Einstein convention for repeated co- and contra-variant indices is adopted; $\lambda$ and $\mu$ are the \textit{Lam\'{e} coefficients} of the medium. The term $\epsilon_i^i=\epsilon$ is the invariant trace of the strain tensor; $\epsilon_{ij}\epsilon^{ij}$ is the invariant trace of $\mathbf{\epsilon} \otimes \mathbf{\epsilon}$. In principle the series goes on to higher order terms, calling in higher order scalars of the strain tensor and more parameters, however we shall stay with the linear approximation elasticity. The actual elastic behavior is insured by the fact that any strain is associated with a corresponding stress, $\sigma_{ij}$, and the (linear) relation among strains and stresses is given by Hooke's law (in tensorial form):
\begin{equation}
\sigma_{ij}=C_{ij}^{kl}\epsilon_{kl}.
\label{hooke}
\end{equation}
$C_{ijkl}$ is the rank-4 elastic moduli tensor, containing the properties of the medium. Its explicit form is given in terms of the unstrained metric tensor:
\begin{equation}
C_{ijkl}=\lambda E_{ij}E_{kl}+\mu(E_{ik}E_{jl}+E_{il}E_{jk}).
\label{moduli}
\end{equation}
The elastic deformation energy density turns out to be:
\begin{equation}
W=\frac{1}{2}\sigma_{ij}\epsilon^{ij}=\frac{1}{2}\lambda \epsilon^2+\mu\epsilon_{ij}\epsilon^{ji}.
\label{elastic}
\end{equation}

Now all tools are ready for the application of the theory to space-time.

\section{The 'Elastic' Space-Time}

Using the results of the previous section and the classical form for the action integral we may write:
\begin{equation}
S=\int{[ R-\frac{1}{2}(\lambda\epsilon^2+2\mu\epsilon_{\alpha\beta}\epsilon^{\beta\alpha})+2\kappa\mathfrak{L}_{mat}]\sqrt{-g}d^4x}.
\label{elas}
\end{equation}
$R$ is the curvature scalar and plays the role of kinetic term, since it contains the derivatives of the elements of the metric tensor; then we have the 'elastic' terms, acting as a potential energy density; $\mathfrak{L}_{mat}$ is the Lagrangian density of matter; $g$ is the determinant of the metric tensor and is used to express the invariant four-volume element, once the coordinates have been chosen. The matter Lagrangian density is coupled to geometry through the $\kappa$ constant. The "elastic" contributions are actually part of the geometry; instead of a coupling constant they contain the properties of the physical space-time expressed by the Lam\'{e} coefficients $\lambda$ and $\mu$. A little notation change has been introduced in order to comply with the GR conventions: Greek letters for the indices run from 0 (time) to 3; Latin indices label space variables only and range from 1 to 3.

Isolating the terms expressing the only properties of space-time we have the strained state Lagrangian density:
\begin{equation}
\mathfrak{L}=(R-\frac{1}{2}\lambda\epsilon^2-\mu\epsilon_{\alpha\beta}\epsilon^{\beta\alpha})\sqrt{-g}.
\label{ssl}
\end{equation}

From (\ref{ssl}) we may build an effective canonical energy/momentum tensor, that acts as a source in addition to the one originated from matter distributions. The elements of this 'elastic' energy/momentum tensor are:
\begin{equation}
T_{(e)\mu\nu}=\lambda\epsilon(\epsilon_{\mu\nu}-\frac{1}{4}\epsilon g_{\mu\nu})\sqrt{-g}+
2\mu(\epsilon_{\mu\alpha}\epsilon^\alpha_\nu-\frac{1}{4}g_{\mu\nu}\epsilon_{\alpha\beta}\epsilon^{\beta\alpha})\sqrt{-g}.
\label{tela}
\end{equation}

It is immediately seen that the trace, $T_{(e)\mu}^{\mu}$ of the tensor (\ref{tela}) is identically zero, so that in vacuo the Einstein equations reduce to:
\begin{equation}
R_{\mu\nu}=-\frac{1}{2}T_{(e)\mu\nu}.
\label{elasticEinstein}
\end{equation}
In practice the energy distribution of the strain fluid looks like the one of a pure radiation field.

\subsection{Robertson-Walker symmetry}

\begin{figure}[tbp]
\vspace{2cm}
\centerline{\includegraphics[width=9cm]{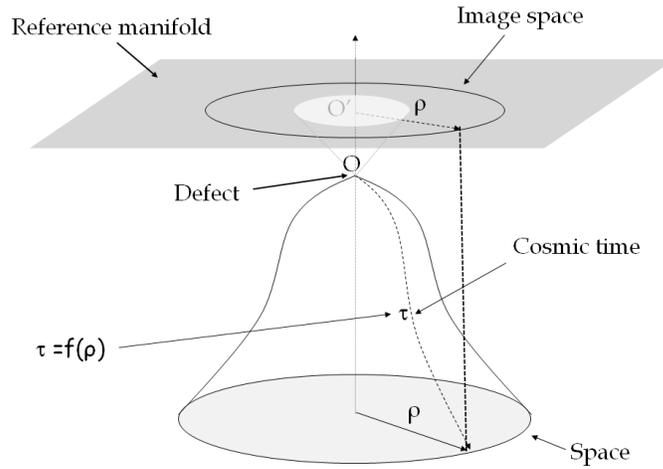}}
\vspace{0cm}
\caption{Pictorial view of a Robertson-Walker space-time. The grey flat area above represents the unstrained frame from which the actual manifold comes, after cutting away the white circle and shrinking the rim to a point in \textit{O}. \textit{O} is a texture defect that induces the bell-shaped manifold representing the actual space-time. Successive sections perpendicular to the axis identify the expanding space and make the accelerated trend manifest after an initial decelerated expansion phase. \label{f1}}
\end{figure}

The prevailing opinion is that the universe has globally a Robertson-Walker symmetry, i.e. it is homogeneous and isotropic in space. This idea is more or less corroborated by observation at the cosmic scale, but is not a necessary consequence of the hypothesis that matter behaves (again at the appropriate scale) as dust. In the Strained State Cosmology (SSC) the global symmetry may be motivated as being due to the presence of a symmetry-fixing cosmic defect. This view may be better understood looking at Fig.~\ref{f1}.

The logical sequence runs as follows. Start from a Minkowskian (or Euclidean) unstrained manifold (the grey surface in the upper part of the figure). Then cut out a circle (the white area in the figure) and pull the rim inward untill the hole reduces to the point \textit{O}. Treating the manifold as a material continuum, the above process generates a defect, according to the definition given by Vito Volterra \cite{volterra}; the defect entails a spontaneous strain in the continuum. Ideally embedding our manifold in a higher dimensional flat space, the strained state appears as a peculiar deformation, which, in the example, has the shape of a bell.\cite{cqg} The cosmic time $\tau$ is measured along geodetics of the surface, starting from the singularity (the defect) in the origin. Space is represented by orthogonal sections of the figure, appearing as circles (actually they would be spheres of simultaneous events in the reference frame of an observer co-moving with the cosmic fluid). The variable $\rho$ is not a distance, but the curvature radius of space. Visibly the sequence of the sections from the singularity towards infinity grow with an initial decelerated, then accelerated trend. The universe sketched in Fig.~\ref{f1} is closed, because that is the simplest situation to be drawn. In the real universe, space seems to be flat, which condition is easily reproducible substituting a linear defect for the point-like one used in the figure.

Coming to the formulae, we write the line element of the Robertson-Walker space-time (the natural manifold) in the traditional way:
\begin{equation}
ds^2=d\tau^2-a^2(\tau)dl^2,
\label{dsRW}
\end{equation}
where $a(\tau)$ is the scale factor of the universe (depending on the cosmic time only) and $dl$ is the space line element.\footnote{
I am directly considering a flat space.}

The above line element must be compared with the corresponding one on the reference unstrained manifold (which no longer exists, but is used to fix the logic just as the embedding does). Using the same coordinates it is:
\begin{equation}
ds_{ref}=b^2(\tau)d\tau^2+kdl^2.
\label{dsref}
\end{equation}
Here $b(\tau)$ is a gauge function accounting for the size of the excluded region that will give rise to the defect. In practice $b$, depending on the $\tau$ coordinate only, fixes the one-to-one correspondence between points on the natural and on the reference manifold (vertical line in the Fig.~\ref{f1}, drawn before the generation of the defect). The parameter $k$ is $+1$ if we assume the reference manifold to be Euclidean, $-1$ if the unstrained manifold is assumed to be Minkowskian. The physical conclusions will be the same; the formal aspects related to the choice of the reference signature are discussed in Ref.~\refcite{signature}.

Using (\ref{dsRW}) and (\ref{dsref}) we may immediately write the elements of the strain tensor defined in (\ref{strain}). The non-zero elements are:
\begin{equation}	
\epsilon_{00}=\frac{1-b^2}{2}, \hspace{8pt} \epsilon_{ii}=-\frac{a^2
+k}{2}.
\label{epsRW}
\end{equation}

Introducing the elements (\ref{epsRW}) into Eq.~(\ref{ssl}) we get, after an integration by parts in order to reduce a linear second order time derivative, the Lagrangian density of our Robertson-Walker manifold:
\begin{equation}
\mathfrak{L}=6a\dot{a}^2-\lambda\frac{4a^2-a^2b^2+3k}{8a}-\mu\frac{a^4(b^2-1)^2+3(a^2+k)^2}{4a}.
\label{lagrRW}
\end{equation}

The Lagrangian does not contain any derivative of the gauge function $b$, so it is a trivial matter to deduce $b$ as a function of $a$:
\begin{equation}
b^2=2\frac{B}{\mu}+3\frac{k}{a^2}\frac{\lambda}{\lambda+2\mu},
\label{b2}
\end{equation}
where $B$ is an abbreviation: $B=\mu\frac{2\lambda+\mu}{\lambda+2\mu}$.

Finally the Lagrangian density of the strain field, in terms of the only scale factor, becomes:
\begin{equation}
\mathfrak{L}=6a\dot{a}-\frac{3}{4}\frac{\lambda\mu}{\lambda+2\mu}\frac{a^2+k}{a}-3\mu\frac{\lambda^2+\lambda\mu+\mu^2}{(\lambda+2\mu)^2}\frac{(a^2+k)^2}{a}.
\label{lagrel}
\end{equation}

\subsection{The effect of matter}

So far the only empty space-time has been considered, evidencing the effect of a cosmic texture defect. Adding matter, in the form of dust and radiation, does not change the symmetry determined by the cosmic defect; rather the dust complies with the given symmetry. Adding to (\ref{lagrel}) the terms for dust and radiation we describe a Friedmann-Lema\^{\i}tre-Robertson-Walker (FLRW) universe where also the cosmic strain is accounted for. Proceeding in the standard way and applying the energy condition it is possible to obtain the Hubble parameter, $H$, for our universe:\cite{mnras}
\begin{equation}
H=\frac{\dot{a}}{a}=c\Big\{\frac{\kappa}{3}(1+z)^3[\rho_{m0}+\rho_{r0}(1+z)]-\frac{B}{4}\Big(1+\frac{(1+z)^2}{a_0^2}\Big)^2\Big\}^{1/2}.
\label{hubble}
\end{equation}
The new symbols appearing in (\ref{hubble}) are: $z$ representing the cosmic redshift parameter; $\rho_{m0}$, the present matter density in the universe; $\rho_{r0}$, the present radiation energy density; $a_0$, the present value of the scale factor.

From Eq.~(\ref{lagrel}) it is also possible to deduce the energy density, $\rho_ec^2$, of the strain field and the ensuing pressure, $p_e$:
\begin{equation}
\rho_ec^2=-\frac{3}{4}B\frac{(a^2+k)^2}{a^4}, \hspace{8pt} p_e=\frac{B}{4}\frac{3a^4+2ka^2-1}{a^4}.
\label{stfluid}
\end{equation}

From (\ref{stfluid}) the equation of state parameter $w$ (from $p_e=w\rho_ec^2$) follows:
\begin{equation}
w=-\frac{3a^4+2ka^2-1}{3(a^2
+k)^2}.
\label{w}
\end{equation}

It is easily seen that:
\begin{equation}
w_{a\rightarrow 0}=\frac{1}{3}, \hspace{8pt} w_{a\rightarrow \infty}=-1;
\label{limits}
\end{equation}
which means that the strain filed behaves like radiation in the vicinity of the defect (the initial singularity) and mimics the cosmological constant in the asymptotic regions, for large scale parameters.

\section{Cosmological Tests}

The SSC theory has been subjected to some typical cosmological tests, in order to verify Eq. (\ref{hubble}) and determine the optimal value of the specific parameter, $B$, of the theory (besides the others which are in common with other theories). The tests concerned the dependence of the luminosity of type Ia supernovae on $z$; the primordial nucleosynthesis; the CMB acoustic horizon; the Barionic Acoustic Oscillation; the structure formation after the recombination era. The results and all technical details are published in Ref.~\refcite{mnras} and show the theory being as viable as the standard $\Lambda CDM$. Here I recall, in short, the optimal values of the parameters:
\begin{eqnarray}
B &=& (2.28\pm 0.08)\times 10^{-52}\hspace{3pt}m^{-2}, \nonumber \\[8pt]
\rho_{m0} &=& (2.45\pm 0.15)\times 10^{-27} \hspace{3pt}kg/m^3, \nonumber \\[8pt]
B_{a_0}^{-1} &=& (0.012\pm 0.06)\times 10^{52}\hspace{3pt}m^{2}.
\label{values}
\end{eqnarray}

$B_{a_0}$ includes two of the parameters in (\ref{hubble}), $\rho_{r0}$ and $a_0$:
\begin{equation}
B_{a_0}=\frac{8}{9}\kappa\rho_{r0} a_0^4.
\end{equation}

\section{SSC versus Massive Gravity}

The Lagrangian density of the Strained State Theory (SST) resembles the one introduced by Fierz and Pauli in 1939 \cite{FP} with the purpose of studying the possibility that gravitons (provided General Relativity is quantizable) have a mass $m$:
\begin{equation}
\mathfrak{L}_{FP}=\frac{m^2}{4}(h^2-h_{\alpha\beta}h^{\beta\alpha}).
\label{FiPa}
\end{equation}
The $h$s appearing in (\ref{FiPa}) are first order differences between the actual metric $g_{\mu\nu}$ and the Minkowski metric $\eta_{\mu\nu}$. Formally the SST Lagrangian goes over the Fierz and Pauli Lagrangian if $\frac{\lambda}{2}=-\mu=m^2$.
Fierz and Pauli's theory suffers the so called van Dam-Veltman-Zakharov discontinuity\cite{vdv,Za}: the solutions found with a non-zero mass of the graviton do not reproduce the corresponding GR solutions when $m \rightarrow 0$.

One more trouble with the Lagrangian (\ref{FiPa}) is that, in the framework of quantum field theory, it produces {\it ghosts}. This problem has been discussed during the 70s of the past century and various mechanisms have been devised to eliminate the ghosts (see references cited in Ref.~\refcite{gaba}).

Without entering into details, I just recall the fact that the Fierz and Pauli Lagrangian (as well as others proposed in order to develop massive gravity theories) is a linearized approximation of the real problem, whereas the SST is 'exact', i.e. the strain tensor $\epsilon_{\mu\nu}$ is proportional to the full difference between the real metric and the Euclidean (or Minkowskian) metric of the flat reference manifold. There is no problem with the non-uniform convergence of a series.\footnote{The various mechanisms proposed to cure the problem of ghosts are based on cancellations produced by higher order terms of the series with respect to the lower order ones.} By the way the  cosmological solution of the SST smoothly reproduces a FLRW universe when $\lambda,\mu,$ then $B$ go to zero.

Some non-linear massive gravity theories introduce an auxiliary metric that enters an effective Lagrangian looking like the Fierz and Pauli's one. It is important to stress that the SST is no bi-metric theory: the only real metric is $g_{\mu\nu}$. The $E_{\mu\nu}$ appearing in the definition of the strain tensor is, in a sense, a logic tool, but has no role as a metric tensor.

Summing up, apparently the various inconveniences that affect the massive gravity theories do not harm the SST, which (it is important to highlight this point) is an entirely classical theory.

\section{Open Questions}

In our ordinary three-dimensional world, we know that the elasticity of a continuum is an emerging property of the underlying interactions among the elementary constituents of the material under consideration. Should we expect to be like that also in the case of space-time? Maybe. In the past there have been attempts to ascribe the 'rigidity' of space-time to the zero-point energy of the quantum vacuum.\cite{sacha} These approaches resemble more to higher order (in $R$) theories rather than to the SST, but in any case they had to face the problem of the many orders of magnitude mismatch between the energy density of the quantum vacuum and the one attributable to the curvature.

The problem is important, but, as far as the grand scale is concerned, it does not affect the results. It would become dominant and require a solution that nobody has so far been able to find, at the highest energy densities and shortest distances, where the conflict between Quantum Mechanics and General Relativity becomes manifest.

Another delicate and general question concerns the nature and role of space-time, on one side, and matter, on the other. Is the universe really dual? Could one of the two ingredients be reduced to the other? A suggestive idea is that of getting rid of matter re-interpreting it in terms of defects of the four-dimensional continuum. In other words, instead of having just one cosmic defect, could the matter be described as a host of soldered cracks in the four-dimensional manifold? Each crack, viewed as a linear (not straight) time-like defect, could induce a local strain which we normally read as a gravitational field. This issue is entirely open and for the moment the vague idea of matter as a network of defects is not much more than hand waving, but who knows?

\section{Conclusion}

Assuming that space-time is real and endowed with physical properties analogous to the ones we know with ordinary three-dimensional deformable materials, we have seen that gravity may be interpreted as a consequence of the strain of the manifold. Applying this idea to the whole universe we have seen that the global Robertson-Walker symmetry can be induced by a cosmic texture defect corresponding to the initial singularity of the traditional GR cosmology; the singularity, in this case appears a bit more understandable and milder than in the usual Big Bang approach. The theory, at the cosmic scale, has just one parameter (in general they are two) and is able to positively reproduce various features of the observed universe, first of all the accelerated expansion. The behavior of the strained space-time manifold close to the cosmic defect even mimics the exponential expansion, which is normally described in terms of inflation. The strained state with its strain energy density provides also a simple and intuitive interpretation of the dark energy.

All in all the SST looks promising and appealing, though its formulation in terms of elementary interactions has not been attempted yet.

\end{document}